\documentclass[nofootinbib,twocolumn,prd,superscriptaddress,preprintnumbers]{revtex4}

\usepackage{graphicx}				
				
\usepackage{amssymb}
\usepackage{amsmath}
\usepackage{mathrsfs}
\usepackage{bbm}
\usepackage{color}
\usepackage{verbatim}
\usepackage{bm}

\def\bea{\begin{align}}
\def\eea{\end{align}}

\newcommand{\mfi}{{\mathfrak{i}}}
\newcommand{\mfj}{{\mathfrak{j}}}
\newcommand{\mfk}{{\mathfrak{k}}}
\newcommand{\mfl}{{\mathfrak{l}}}
\def\vev#1{\left\langle #1 \right\rangle}
\def\nn{\nonumber \\ }
\def\rd{ {\rm d} }
\def\McD{ {\mathcal{D} } }
\def\MsD{ {\mathscr{D} } }

\usepackage{mathrsfs}

\begin{document}

\preprint{CERN-PH-TH-2015-257}

\title{A Geometric Formulation of Higgs Effective Field Theory: Measuring the Curvature of Scalar Field Space}

\author{Rodrigo Alonso}

\affiliation{\vspace{1mm}
Department of Physics, University of California at San Diego, La Jolla, CA 92093, USA}

\author{Elizabeth E.~Jenkins}
\author{Aneesh V.~Manohar}

\affiliation{\vspace{1mm}
Department of Physics, University of California at San Diego, La Jolla, CA 92093, USA}

\affiliation{\vspace{1mm} CERN TH Division, CH-1211 Geneva 23, Switzerland}


\begin{abstract}
A geometric formulation of Higgs Effective Field Theory (HEFT) is presented.   Experimental observables are given in terms of geometric invariants of the scalar sigma model sector such as the curvature of the scalar field manifold $\mathcal{M}$. We show how the curvature can be measured experimentally via Higgs cross-sections, $W_L$ scattering, and the $S$ parameter. The one-loop action of HEFT is given in terms of geometric invariants of $\mathcal{M}$. The distinction between the Standard Model (SM) and HEFT is whether  $\mathcal{M}$ is flat or curved, and the curvature is a signal of the scale of new physics.
\end{abstract}
\maketitle

\section{Introduction}

The recent discovery of a neutral scalar particle with a mass of $\sim 125$\,GeV has renewed interest in non-linear effective Lagrangians.  An effective Lagrangian for a spontaneously broken gauge theory with three ``eaten" Goldstone bosons and one additional neutral scalar particle yields the most general low-energy description of the interactions of the new scalar particle detected at the LHC.  
This Higgs Effective Field Theory (HEFT) Lagrangian~\cite{Feruglio:1992wf,Grinstein:2007iv} is constructed with the three ``eaten'' Goldstone bosons and the light neutral Higgs boson transforming as a triplet and a singlet, respectively, under the custodial symmetry.  
The only implicit assumption in this HEFT framework is that there are no other light states in the few hundred GeV range which couple to SM particles, which is known to be satisfied experimentally.  
The theory contains as a limiting case the renormalizable Standard Model (SM) Higgs Lagrangian, where the neutral scalar and the Goldstone bosons of the spontaneously broken electroweak gauge symmetry form a complex scalar doublet $H$ that transforms linearly as $\mathbf{2}_{1/2}$ under the electroweak gauge symmetry $SU(2)_L \times U(1)_Y$.  
An important special case of HEFT is the Standard Model Effective Field Theory (SMEFT), where the scalar fields of the EFT transform linearly as a complex scalar doublet $H$.  Schematically,
\begin{eqnarray}
{\rm SM} \subset {\rm SMEFT} \subset {\rm HEFT}.
\end{eqnarray}

The path integral formulation of quantum field theory gives a prescription for computing the $S$-matrix of the theory from the Lagrangian.  An important result in quantum field theory is that the $S$-matrix is independent of the fields chosen to parametrize the theory: field redefinitions which change the form of the Lagrangian leave the $S$-matrix invariant.  The well-known analysis of sigma models by Callan, Coleman, Wess and Zumino (CCWZ)~\cite{Coleman:1969sm,Callan:1969sn} uses this freedom to make field redefinitions on the scalar fields to put all spontaneously broken theories into a standard form.  The CCWZ convention specifies a definite choice for the scalar fields of a given sigma model.  Although making such a choice eliminates the ambiguity of the Lagrangian due to field redefinitions, it does so at a cost of obscuring the geometry of the sigma model, which describes the field-independent properties.  In this work, we present a geometric formulation of HEFT which emphasizes the field-independent observables of the scalar sigma model sector.
This geometric formulation of HEFT makes explicit the connection between the geometry of scalar field space and experimental measurements.

There are many important features of the scalar sector of a spontaneously broken theory that are obscured in the standard presentation of the Higgs sector of the SM.  
The usual SM Higgs sector with a fundamental Higgs doublet can be written in the universal CCWZ formulation as a non-linear effective Lagrangian of a Goldstone boson triplet and an additional neutral  singlet scalar field $h$. The non-linear version leads to the \emph{same} $S$-matrix as the linear formulation, including quantum corrections.  It is clear that adding the scalar singlet $h$ with precisely the right couplings is key to the equivalence of the non-linear and linear parameterizations of the SM Higgs sector.  The non-linear parameterization of the SM Higgs sector in which the field $h$ is not 
necessarily part of a doublet can be consistently generalized to a non-renormalizable EFT, which is HEFT.  Theories in which $h$ is the dilaton or a Goldstone boson of an enlarged global symmetry provide examples of HEFT.  
SMEFT is a special case of HEFT since it requires that the scalar $h$ and the three Goldstone bosons transform as a complex scalar doublet.

In the literature, SMEFT and HEFT are referred to as the linear Lagrangian and the non-linear or chiral Lagrangian, respectively.  This nomenclature is somewhat misleading, given that SMEFT is a special case of HEFT.  Throughout this paper, we use the term HEFT to refer to the most general EFT containing the physical Higgs boson.  We focus on theories with custodial $SU(2)$, which have the symmetry breaking pattern $O(4) \to O(3)$.
We use linear and non-linear to refer to whether the scalar fields transform linearly or non-linearly under the $O(4)$ symmetry.

HEFT describes the dynamics of the physical Higgs scalar $h$, and the Goldstone bosons 
$\varphi$ from ${\cal G} \to {\cal H}$ symmetry breaking, which together form coordinates on a scalar manifold $\mathcal{M}$.  
Different parameterizations of the scalar sector correspond to different coordinate choices on $\mathcal{M}$.  The $S$-matrix is unchanged by such scalar field redefinitions, i.e.\ by coordinate transformations in scalar field space, and depends only on the geometry of $\mathcal{M}$.  
From this perspective, the key question is not whether $SU(2)_L \times U(1)_Y$ gauge symmetry is realized linearly or non-linearly at the level of the Lagrangian,\footnote{Note that this commonly used terminology of linear and non-linear electroweak symmetry breaking derives from the non-linear chiral Lagrangian for QCD.}
since one can convert from the linear form to the non-linear form by a field redefinition,
but whether the scalar manifold $\mathcal{M}$ is \emph{curved} or \emph{flat}.  The renormalizable SM has a flat scalar manifold.  A geometric formulation of HEFT makes clear that all physical observables are independent of the parameterization of the scalar fields.   
SMEFT is a special case of HEFT where there is an $O(4)$ invariant point on the scalar manifold.  The SMEFT Lagrangian is given by expanding the HEFT Lagrangian about this special point (which is at $H=0$) in a power series in the scalar fields.

In this paper, we compute the Riemann curvature tensor of $\mathcal{M}$, and show how the curvature can be measured experimentally in terms of the couplings of the physical Higgs boson to the massive electroweak gauge bosons $W^\pm$ and $Z$.  We also explore other couplings of the Higgs boson to the massless SM gauge boson, the photon, and to fermions from a geometric point of view.  We emphasize that an important goal of precision Higgs boson physics will be to constrain the curvature of the scalar manifold $\mathcal{M}$, which is a measure of the scale of new physics.

The geometric formulation of non-linear sigma models is well-known, and has been used extensively for supersymmetric sigma models, and, to a lesser extent, for chiral perturbation theory~\cite{Honerkamp:1971sh,Tataru:1975ys,AlvarezGaume:1981hn,AlvarezGaume:1981hm,Gaillard:1985uh}.  In this paper, we apply it to HEFT with a single light neutral scalar $h$.  A general formulation of spontaneously broken ${\cal G}/{\cal H}$ theories with an arbitrary number of additional scalars is given in a subsequent work~\cite{p2}.

A coordinate-invariant formulation of HEFT also clears up a number of subtleties which arise in the one-loop corrections to the theory.  
Calculations of radiative corrections in sigma models by Appelquist and Bernard~\cite{Appelquist:1980vg,Appelquist:1980ae}, and more recently for HEFT by Gavela {\it et al.}~\cite{Gavela:2014uta}, require intermediate steps in which chiral non-invariant terms proportional to the equations of motion (EOM) are redefined away. The appearance of these terms for curved scalar manifolds does not have physical implications, and they are avoided when employing a covariant formalism for perturbation theory~\cite{Honerkamp:1971sh,Tataru:1975ys}.
In our geometric formulation of HEFT, quantum corrections to the theory are given in terms of the curvature of the scalar manifold 
$\mathcal{M}$, and non-covariant terms which vanish on-shell can be understood and systematically dealt with.

The organization of this paper is as follows.  Section~II defines the curvature of the scalar field manifold for HEFT, and gives ways to measure the curvature experimentally. Section~III presents the path integral formalism for calculating radiative corrections in an arbitrary sigma model.  A geometric formulation of all quantities is given.  Section~IV specializes to renormalization in the case of interest, namely HEFT. Section~V gives the conclusions.

\section{Curvature of  Scalar Field Space\label{sec:curvature}}

The Higgs sector of the SM has a complex scalar doublet $H$ which can be defined in terms of 4 real scalar fields $\phi_H^i$, $i=1,2,3,4$, by
\begin{align}
H &= \frac{1}{\sqrt 2} \left[ \begin{array}{cc} \phi_H^2 + i \phi_H^1 \\ \phi_H^4 - i \phi_H^3 \end{array}\right].
\end{align}
The scalar potential $V(H)$ depends only on the magnitude of the scalar 4-vector
\begin{align}
(\phi_H^1)^2+ (\phi_H^2)^2+ (\phi_H^3)^2+ (\phi_H^4)^2 \equiv 2 H^\dagger H,
\end{align}
and it has a minimum at the vacuum expectation value $v=246$~GeV,
\begin{align}
\langle (\phi_H^1)^2+ (\phi_H^2)^2+ (\phi_H^3)^2+ (\phi_H^4)^2 \rangle &=  v^2,
\end{align}
which spontaneously breaks ${\cal G}=O(4)$ symmetry down to ${\cal H}=O(3)$.  
It is convenient to define the $2 \times 2$ scalar field matrix
\begin{eqnarray}
\Sigma &\equiv& \left(\tilde H\,, H\right) ,
\end{eqnarray}
where $\tilde H \equiv (i \sigma_2) H^*$.
Under the $O(4) \sim SU(2)_L \times SU(2)_R$ symmetry, $\Sigma$ transforms as
\begin{align}\label{tildeHH}
\Sigma  &\to L \Sigma R^\dagger, & \left\langle \Sigma \right\rangle& =\frac{v}{\sqrt2}\mathbbm 1,
\end{align}
where $L$ and $R$ are $2\times 2$ unitary matrices, and $\mathbbm 1$ is the $2 \times 2$ identity matrix.
One sees that the unbroken ${\cal H}$ symmetry is custodial $O(3)\sim SU(2)_V$, which ensures the gauge boson mass relation $M_W=M_Z \cos \theta_W$ at tree level.  The vacuum manifold ${\cal G}/{\cal H}$ is isomorphic to the three-sphere $S^3$ and is parametrized by three Goldstone boson coordinates $\varphi^a$, $a=1,2,3$ where Latin letters from the beginning of the alphabet
will be used to distinguish $O(3)$ indices from $O(4)$ indices, $i=1,2,3,4$.  The radial direction perpendicular to $S^3$ corresponds to the Higgs boson direction $h$, which transforms as a singlet under the unbroken symmetry group. We can make this geometric relationship explicit using spherical polar coordinates in scalar field space:
\begin{align}\label{udef}
\left[ \begin{array}{c} \phi_H^1 \\ \phi_H^2 \\ \phi_H^3 \\\phi_H^4 \end{array} \right] 
&\equiv \left(v+h\right)   \left[ \begin{array}{c} u^1(\varphi) \\ u^2(\varphi) \\ u^3(\varphi) \\  u^4(\varphi) \end{array} \right] , &\begin{array}{c}u(\varphi) \cdot u (\varphi) =1,\\[3mm]
u^i(0)=\delta^{i4},
\end{array}
\end{align}
where $u^i(\varphi)$ is a unit 4-vector depending only on the angular coordinates $\bm{\varphi}$ on $S^3$, and $h$ is the radial coordinate.

The above discussion focuses on the global symmetries of the $O(4)$ sigma model.  Now, we account for the partial gauging of the $O(4)$ symmetry. 
The electroweak gauge symmetry group ${\cal G}_{\rm gauge}$ in the Higgs sector is a subgroup of the global symmetry group, ${\cal G}_{\rm gauge} \subset {\cal G}$.
Specifically, using the definition of $\Sigma$ in Eq.~(\ref{tildeHH}),
$O(4)\sim SU(2)_L\times
SU(2)_R \supset SU(2)_L\times U(1)_Y$, where $U(1)_Y$ is the Abelian gauge symmetry associated with the diagonal generator of $SU(2)_R$. This decomposition is made explicit in the gauge covariant derivative
\begin{widetext}
\begin{align}\label{GgCovDFlat}
D_\mu \phi_H=\partial_\mu \phi_H+i\left(gW_\mu^IT_I+g^\prime B_\mu Y\right)\phi_H=\partial_\mu \phi_H
+\frac12\left( \begin{array}{cccc}
0&gW_\mu^3+g^\prime B_\mu & -gW_\mu^2&gW_\mu^1\\
-gW_\mu^3-g^\prime B_\mu &0& gW_\mu^1&gW_\mu^2\\
gW_\mu^2&-gW_\mu^1&0& gW_\mu^3-g^\prime B_\mu \\
-gW_\mu^1&-gW_\mu^2&-gW_\mu^3 +g^\prime B_\mu&0\\
\end{array}\right) \phi_H.
\end{align}
\end{widetext}
Note that
\begin{align}D_{\mu}\phi_H^i=(v+h) \left[ D_\mu u^i(\varphi) \right] +u^i(\varphi) \left( \partial_\mu h \right),
\label{Dudef}
\end{align} 
since the gauge group does not act on the radial field $h$.
The gauge kinetic energy term is
\begin{align}
\label{KEHiggs}
{\mathscr L} &= \frac 12 \left( D_\mu \phi_H^i \right) \left(D^\mu \phi_H^i\right)\nn
&=\frac 12 (v+ h)^2 \left( D_\mu u^i(\varphi) \right) \left( D^\mu u^i(\varphi) \right)+ \frac 12 \left( \partial_\mu h \right)^2 \\
&= \frac 12 \left(1+ \frac h v \right)^2 g_{ab} (\varphi )\left( D_\mu \varphi^a \right) \left( D^\mu \varphi^b \right) + \frac 12 \left( \partial_\mu h \right)^2, \nonumber
\end{align}
where
\begin{eqnarray}
D_\mu u^i(\varphi) &=& \left( D_\mu \varphi^a \right) \left(\frac{\partial u^i(\varphi)}{\partial \varphi^a} \right)
\end{eqnarray}
and
\begin{eqnarray}
g_{ab}(\varphi) &\equiv&  v^2 \left(\frac{\partial u^i(\varphi)}{\partial \varphi^a} \right) \left(\frac{\partial u^i(\varphi)}{\partial \varphi^b} \right)
\end{eqnarray}
have been used to obtain the last line.

The three angular Goldstone bosons $\varphi^a$ and the radial Higgs field $h$ define spherical polar coordinates $\phi_H^\mfi \equiv ( \varphi^a, h )$ for the scalar manifold. The gothic index $\mfi= \{a,h\}$ denotes the four spherical polar coordinates for the $O(4)$ sigma model.
In spherical polar coordinates, the scalar kinetic energy term is
\begin{eqnarray}
{\mathscr L} &=& \frac 12 g_{\mfi \mfj}(\phi_H) \left( D_\mu \phi_H^\mfi \right) \left( D^\mu \phi_H^\mfj \right),
\end{eqnarray}
which defines a metric for the scalar manifold ${\cal M}$.  Eq.~(\ref{KEHiggs}) gives the metric of the SM Higgs sector 
\begin{eqnarray}
g_{\mfi \mfj}(\phi_H) &\equiv&  \left[ \begin{array}{cc} (1 + h/v )^2 g_{ab}(\varphi) & 0  \\ 0  & 1  \end{array} \right],
\label{SMmetric}
\end{eqnarray}
where $g_{ab}(\varphi)$ is the $O(3)$ invariant metric on the Goldstone boson coset space ${\cal G}/{\cal H} = S^3$ with radius $v$.

Many different parametrizations of $u^i(\varphi)$ are possible.  For the choice of the square root parametrization, 
the scalar coordinates $\phi_H^i$ are given by 
 \begin{align}
\left[ \begin{array}{c} \phi_H^1 \\ \phi_H^2 \\ \phi_H^3 \\\phi_H^4 \end{array} \right] 
&= \left(1+\frac{h}{v}\right) 
\left[ \begin{array}{c} \varphi^1 \\ \varphi^2 \\ \varphi^3 \\  \sqrt{v^2- \bm{\varphi \cdot \varphi}} \end{array} \right]\ ,
\label{6}
\end{align}
and Eq.~(\ref{KEHiggs}) equals
\begin{align}
{\mathscr L}
&= \frac 12 \left(1+\frac h v \right)^2 \left[ D_\mu \bm{\varphi} \cdot D^\mu \bm{\varphi} + \frac{ (\bm{\varphi} \cdot D_\mu
\bm{\varphi})^2}{v^2- \bm{\varphi \cdot \varphi}} \right]+\frac12 \left(\partial_\mu  h \right)^2 ,
\end{align}
which implies that
\begin{align}
g_{ab}(\varphi) &= \left[ \delta_{ab} +  \frac{ \varphi_a \varphi_b}{v^2 - \bm{\varphi \cdot \varphi}} \right] .
\label{13}
\end{align}

The electroweak gauge symmetry is contained in ${\cal G}$ but only partially in ${\cal H}$; the overlap of
the electroweak gauge symmetry and ${\cal H}$ which remains unbroken is ${\cal H}_{\rm gauge} =U(1)_Q$, the Abelian electromagnetic gauge symmetry.
The three Goldstone boson modes $\varphi^a$ of the ${\cal G}/{\cal H}$ global symmetry breakdown are in one-to-one correspondence with the broken gauge symmetry
generators of ${\cal G}_{\rm gauge}/{\cal H}_{\rm gauge}$, and they become the longitudinal polarization states of the three massive gauge bosons $W^\pm$ and $Z$. The effect of the broken gauge symmetry on $\phi_H^i$ is a {\it local rotation} that only
affects $u^i(\varphi)$; this transformation can be chosen to remove all three $\varphi^a$ degrees of freedom from $\phi_H$ in 
unitary gauge. The resulting Lagrangian has a single Higgs boson $h$ coupled to three massive gauge bosons $W^\pm$ and $Z$.

One does not have to assume that the full $SU(2)_R$ group is contained in  ${\cal G}$.  Instead, one can consider a theory in which only $U(1)_R$ is in $\mathcal{G}$.  In this case, the ${\cal G} \to {\cal H}$ symmetry breaking is $SU(2)_L \times U(1)_R \to U(1)_V$.  The coset space ${\cal G}/{\cal H}$ is still topologically $S^3$ and parametrized by three coordinates $\varphi^a$, but the metric $g_{ab}(\varphi)$ on ${\cal G}/{\cal H} = S^3$ is no longer the $O(3)$ invariant metric.  Under the unbroken global symmetry group ${\cal H} = SO(2)\sim U(1)_V$, the triplet $\varphi^a$ transforms as a doublet and a singlet.  Then, the $\varphi^{1,2}$ and $\varphi^3$ directions are no longer related by symmetry, and one loses the phenomenologically successful gauge boson mass relation $M_W=M_Z \cos \theta_W$.  For this reason, we will assume that $SU(2)_R$ is a global symmetry for the remainder of this paper.

Like the SM Higgs sector, the HEFT scalar sector has four fields, $\varphi^a$ and $h$, where $\varphi^a$ are the angular coordinates of the Goldstone boson coset space 
${\cal G}/{\cal H} = O(4)/O(3)=S^3$, and $h$ is the radial scalar field direction.  These four real scalar fields together are the spherical polar coordinates 
$\phi^\mfi=\{\varphi^a,h\}$ of a curved scalar manifold $\mathcal{M}$, where $\mfi=\{a,h\}$.  
The kinetic energy term of HEFT is given by
\begin{align}
{\mathscr L} &= \frac 12 g_{\mfi\mfj}(\phi)\, D_\mu \phi^{\mfi} D^\mu \phi^{\mfj},
\label{5}
\end{align} 
where $g_{\mfi\mfj}(\phi)$ is a general metric on $\mathcal{M}$.  The metric $g_{\mfi\mfj}$ breaks into $g_{ab}$, $g_{ah}$, $g_{ha}$ and $g_{hh}$.
There are no $g_{h a}$ and $g_{a h}$ off-diagonal metric terms, because one cannot construct a ${\cal H}$-invariant tensor that mixes the $SO(3)$ singlet $h$ field with the $SO(3)$ triplet $\bm{\varphi}$.  
In addition, a field redefinition of the radial coordinate $h$ allows one to set $g_{hh}=1$. 
Consequently, the metric of HEFT is a simple generalization of Eq. (\ref{SMmetric}),
\begin{align}
g_{\mfi\mfj}(\phi)  &= \left[ \begin{array}{cc} F(h)^2 g_{ab}(\varphi) & 0  \\ 0  & 1  \end{array}
\right],
\label{GMtrc}
\end{align}
where $F(h)$ is an arbitrary function of $h$ normalized such that $F(0)=1$ and $g_{ab}(\varphi)$ is the $O(3)$ invariant metric on the scalar submanifold ${\cal G}/{\cal H} =S^3$. 
In a more general sigma model with more than one singlet field $h$, the $g_{hh}=1$ metric is replaced by a non-trivial radial scalar field metric; this case will be discussed further in Ref.~\cite{p2}.  
An explicit coordinate choice for $g_{ab}(\varphi)$ is given by the square root parametrization of Eq.~(\ref{13}).  However, it is better to think more abstractly in terms of arbitrary coordinates $\varphi$ and the corresponding metric $g_{ab}(\varphi)$ on $S^3$.  The radius of $S^3$ is $v$, which is fixed experimentally to be $v \sim 246$\,GeV from the $W,Z$ masses.

The covariant derivative of the spherical polar scalar coordinates $\phi^\mfi$ 
can be derived using 
\begin{align}
(D_\mu\phi)^\mfi \left( \frac{\partial \phi^i}{\partial \phi^\mfi } \right)= \left(D_\mu \phi^i \right) .
\end{align}
The covariant derivative is
\begin{align}\label{GgCovCurv}
D_\mu \phi^\mfi=\partial_\mu \phi^\mfi+i gW_\mu^I t_I^\mfi(\phi)+i g^\prime B_\mu t^\mfi_Y(\phi) ,
\end{align}
where
\begin{align}\label{GenCurv}
t_I^\mfi (\phi)=&g^{\mfi\mfj}\frac{\partial \phi}{\partial \phi^\mfj} \cdot T_I \cdot \phi, &
t_Y^\mfi (\phi)=&g^{\mfi\mfj}\frac{\partial \phi}{\partial \phi^\mfj} \cdot T_Y\cdot \phi,
\end{align}
are Killing vectors of the gauge symmetry on the scalar manifold.
The anti-symmetry of $T_I,T_Y$ implies
$D_\mu \phi^\mfi=\left( (D_\mu\varphi)^a\,,\partial_\mu h\right)$.
The reason for this simplification is that the symmetry group ${\cal G}$ maps points of coset space ${\cal G}/{\cal H}$ to other points of coset space ${\cal G}/{\cal H}$
without affecting $h$, so that the coset coordinates $\varphi^a$ transform under the ${\cal G}$ action, whereas $h$ remains invariant.

The Riemann curvature tensor, Ricci tensor, and Ricci scalar of the ${\cal G}/{\cal H} = S^3$ submanifold are obtained from the metric Eq.~(\ref{GMtrc}) by restricting to $h=0$.  The non-zero curvature components are
\begin{align}
\widehat R{}_{abcd}(\varphi) &= \frac{1}{v^2} \left(g_{ac} (\varphi)g_{bd}(\varphi)-g_{ad}(\varphi) g_{bc}(\varphi) \right), \nn
\widehat R_{bd}(\varphi) &= \frac{1}{v^2} (N_\varphi-1) g_{bd}(\varphi)= \frac{2}{v^2} g_{bd}(\varphi), \nn
\widehat R &=  \frac{1}{v^2} N_\varphi (N_\varphi-1) = \frac{6}{v^2},
\label{10}
\end{align}
where $N_\varphi=3$ is the number of Goldstone bosons, and we remind the reader that indices $a,b,c,d$ take the values $1,2,3$.  The curvature of $S^3$ is set by the electroweak gauge symmetry breaking scale $v$.  The form Eq.~(\ref{10}) for the Riemann tensor holds because $S^3$ is a maximally symmetric space.

The non-zero components of the Riemann curvature tensor $R_{\mfi\mfj\mfk\mfl}(\phi)$ of the full scalar manifold $\mathcal{M}$ for scalar fields $\phi^\mfi$ (splitting the coordinates $\phi^\mfi$ on $\mathcal{M}$ into $\varphi^a$, $a=1,2,3$, in the three angular directions and $h$ in the radial direction) are 
\begin{align}
R_{abcd}(\phi) &= \left[\frac{1}{v^2}- (F^\prime(h))^2\right] F(h)^2 \left(g_{ac} g_{bd}-g_{ad} g_{bc}\right), \nn
R_{ah b h}(\phi) &=  - F(h) F^{\prime \prime} (h) g_{ab},
\label{11}
\end{align}
and components related to these by the permutation symmetry of the Riemann tensor. $R_{abcd}(\phi)$ is proportional to 
$(g_{ac} g_{bd}-g_{ad} g_{bc})$ because $S^3$ is a maximally symmetric space.

The Ricci tensor $R_{\mfj\mfl}(\phi)$ is 
\begin{align}
R_{bd}(\phi) &=\left\{\left[\frac{1}{v^2}- (F^\prime(h))^2\right] (N_\varphi-1)  -  F^{\prime \prime}(h) F(h) \right\} g_{bd},   \nn
R_{hh}(\phi) & = -  \frac{N_\varphi F^{\prime \prime} (h)}{F(h)}  ,
\label{12}
\end{align}
and the Ricci scalar $R(h)$, or the scalar curvature (pun intended), is
\begin{align}
R(h) &= \left[\frac{1}{v^2}- (F^\prime(h))^2\right] \frac{ N_\varphi (N_\varphi-1)}{F(h)^2} - \frac{2 N_\varphi F^{\prime\prime}(h)}{F(h)}.
\label{13a}
\end{align}
The curvature tensors in Eq.~(\ref{10}) of the ${\cal G}/{\cal H}$ submanifold can be obtained from those of the full scalar manifold $\mathcal{M}$ by setting $F(h)=1$ and restricting the tensor indices to be tangent to the submanifold.

One can define dimensionless radial functions $\mathfrak{R}_{h,4,2,0}(h)$ by
\begin{align}
R_{abcd} &= 
\mathfrak{R}_4(h) \widehat  R_{abcd},  & R_{ahbh} &= 
\mathfrak{R}_{2h}(h) \widehat  R_{ab} \nn
R_{bd} &= 
 \mathfrak{R}_2(h) \widehat  R_{bd}  , &
R_{hh} &= 
\mathfrak{R}_{0h}(h) \widehat R,  \nn
R &= 
\mathfrak{R}_0(h) \widehat  R , 
\end{align}
when ${\cal G}/{\cal H}$ is a symmetric space, as for $O(4)/O(3)$.
From Eqs.~(\ref{11}), (\ref{12}) and~(\ref{13a}),
\begin{align}
\mathfrak{R}_4(h) &= \left[1 - v^2 (F^\prime(h))^2\right]F(h)^2, \nn
\mathfrak{R}_{2h}(h) &= - \frac{v^2 F(h) F^{\prime \prime} (h)}{N_\varphi-1}  \nn
\mathfrak{R}_2(h) &= \frac{\mathfrak{R}_4(h)}{F(h)^2}  +  \mathfrak{R}_{2h}(h), \nn
\mathfrak{R}_{0h} (h) &= \frac{\mathfrak{R}_{2h}(h)}{F(h)^2}, \nn
\mathfrak{R}_0(h)  &= \frac{\mathfrak{R}_4(h)}{F(h)^4} +  \frac{2 \mathfrak{R}_{2h}(h)}{F(h)^2},
\end{align}
with  $F(h)^4\, \mathfrak{R}_0(h) + \mathfrak{R}_4(h) = 2F(h)^2\, \mathfrak{R}_2(h)$.
If the HEFT arises from symmetry breaking based on  compact Lie groups as in composite Higgs models~\cite{Dugan:1984hq,Kaplan:1983fs}, the sectional curvatures are non-negative, so that $\mathfrak{R}_{4}(h) \ge 0$, $\mathfrak{R}_{2h}(h) \ge 0$.

Consider first the SM Higgs kinetic energy term, with
\begin{align}
F(h) = 1 + \frac{h}{v}\,.
\label{14}
\end{align}
One sees immediately from Eqs.~(\ref{10}--\ref{13a}), that even though the Goldstone boson submanifold $S^3$ is curved and has non-vanishing curvature $R_{abcd}(\varphi)$, the full scalar manifold $\mathcal{M}$ is \emph{flat} and has vanishing curvature tensor $R^\mfi{}_{\mfj\mfk\mfl}(\phi)$.  This result should be familiar from ordinary three-dimensional space. One can use spherical polar coordinates in $\mathbb{R}^3$.  Spherical shells, which are surfaces of constant radius, are curved, but the full space $\mathbb{R}^3$ is flat.
The coordinate-invariant statement of the SM Higgs sector is that the scalar manifold $\mathcal{M}$ is \emph{flat}, not whether the scalar sector is linear versus non-linear, which changes under field redefinitions.

In HEFT, one considers a general radial function
\begin{align}
F(h) &= 1 + c_1 \left(\frac hv\right) + \frac12 c_2  \left(\frac hv\right)^2 + \ldots ,
\label{Fexp}
\end{align}
which is defined by its coefficients $c_n$, for all $n \ge 1$, in a power series expansion in $h/v$.
In this case, defining $\mathfrak{r}_i \equiv \mathfrak{R}_i(0)$, one obtains
\begin{align}
\mathfrak{r}_4 &= 1-c_1^2 , &
\mathfrak{r}_{2h} & = -\frac12 c_2 , \nn
\mathfrak{r}_2 &= \mathfrak{r}_4 + \mathfrak{r}_{2h} & 
\mathfrak{r}_{0h} &= \mathfrak{r}_{2h}  &
\mathfrak{r}_0 &=\mathfrak{r}_4+2\mathfrak{r}_{2h}
\end{align}
with $\mathfrak{r}_0 + \mathfrak{r}_4 = 2 \mathfrak{r}_2$.
The SM with $F(h)$ given by Eq.~(\ref{14}) implies $c_1=1$, and $c_n=0$ for all $n \ge 2$, so  all $\mathfrak{r}_{i}= 0$.

In theories where the whole Higgs doublet arises as a Goldstone boson multiplet~\cite{Kaplan:1983fs,Dugan:1984hq} with a symmetry breaking scale $f$, one can identify the
dimensionless curvature with the ratio of the electroweak gauge symmetry breaking vacuum expectation value $v$ and the Goldstone boson symmetry breaking scale $f$:
\begin{align}
\mathfrak{r}_{i}\sim \frac{v^2}{f^2}\equiv \xi\label{xidef},
\end{align}
so that deviations from the SM are controlled by the ratio of the radius of the $S^3$ sphere of
the ${\bm \varphi}$ Goldstone bosons to the effective radius of the manifold $\cal M$ that contains the Higgs singlet as well.
Measuring the components of the curvature tensor $R^{\mfi}{}_{\mfj\mfk\mfl}(\phi)$ directly determines the scale of new physics.

The coefficients $c_{1,2}$, and hence the curvatures, are experimentally measurable.
In unitary gauge, the scalar kinetic energy Lagrangian of Eq.~(\ref{5}) is
\begin{align}
\mathscr L &= \frac12 \partial_\mu  h \partial^\mu h  + \frac{v^2}{8} F(h)^2 \left[ 2 g^2 W^+_\mu {W^-}^\mu + (g^2+{g^\prime}^2) Z_\mu Z^\mu \right] \nn
&=  \frac12 \partial_\mu  h \partial^\mu h  +  \left[1+ 2c_1 \frac{h}{v} + (c_1^2+c_2) \frac{h^2}{v^2} + \ldots \right] \nn
&\times \left[ M_W^2 W^+_\mu {W^-}^\mu + \frac 12 M_Z^2 Z_\mu Z^\mu \right] .
\label{19}
\end{align}
This equation shows that $v \sim 246$\,GeV is fixed by the gauge boson masses, and that $c_1$  and $c_2$ are the $hVV$ and $hhVV$ couplings, respectively, in units of the SM strength, where $VV= W^+W^-$ or $ZZ$.  Precision measurements of $hWW$ and $hZZ$ can determine $c_1$, which also enters electroweak precision observables at one loop. In particular, one can estimate the contribution to the
$S$ parameter \cite{Peskin:1990zt}
\begin{align}
\Delta S=\frac{1}{12\pi}\mathfrak{r}_{4}\log\left(\frac{\Lambda^2}{M_Z^2}\right)\,.
\end{align}
Measuring $c_2$ is much more difficult, because it involves two Higgs bosons.  Both $c_1$ and $c_2$ enter the expressions for the curvature because the curvature tensor depends on the metric up to second derivatives. Higher order coefficients $c_n$, $n \ge 3$, in the expansion Eq.~(\ref{Fexp}) of $F(h)$ only enter in derivatives of the curvature. To test the SM requires determining whether $\mathcal{M}$ is flat or curved, which one can do by measuring $\mathfrak{r}_4$, $\mathfrak{r}_2$ and $\mathfrak{r}_0$. 
The advantage of formulating the constraints in terms of the curvature is that it gives a geometric interpretation of the results, independent of the particular coordinates (choice of scalar fields) used to parameterize the HEFT.

One characteristic of a theory of a set of scalars with a curved manifold $\cal M$ is that, in general, it is not self-consistent to arbitrary high energies. This feature shows up in HEFT, in that the singlet Higgs boson $h$ does not fully unitarize longitudinal gauge boson scattering for
general $c_n$~\cite{Barbieri:2007bh}.
Explicitly, the scattering amplitude of longitudinal $W$-bosons $W_L$ depends on the curvature:
\begin{align}
\mathcal{A}\left( W_L W_L\to W_L W_L \right)=\frac{s+t}{v^2}\mathfrak{r}_4\, ,\nn
\mathcal{A}\left( W_L W_L\to h h \right)= -\frac{2s}{v^2}\mathfrak{r}_{2h}\,.
\label{unitary}
\end{align}
For composite Higgs theories based on compact groups, $\mathfrak{r}_4\ge0$, $\mathfrak{r}_{2h}\ge 0$.
Eqs.~(\ref{unitary}) imply that the more curved the scalar manifold space ${\cal M}$ is, the lighter the required resonances are to fully unitarize the scattering amplitudes.  The scale of new physics governing the mass of these resonances is
$\Lambda\sim 4\pi v/\sqrt{\mathfrak{r}}$~\cite{Manohar:1983md}. Note that this result is in accordance with the scenario of the Higgs boson as a Goldstone boson~\cite{Dugan:1984hq,Kaplan:1983fs} where resonances are expected at $\Lambda\sim 4\pi f$.

\section{Radiative Corrections}

The $S$-matrix for a scalar theory with action $S[\phi]$ can be computed from the $n$-point $\phi$ Green's functions $G^{(n)}$ which are generated by $W[J]$,
\begin{align}
Z[J]\equiv e^{iW[J]}=
\int D \phi\ \mbox{exp}\left[i\left( S[\phi] +\int \rd x \ J \phi\right)\right]\,.
\label{22}
\end{align}
The Green's functions can be constructed from tree graphs with 1PI vertices $\Gamma^{(n)}$, which, in turn, are generated by $\Gamma[\widetilde \phi]$, the Legendre transform of $W[J]$,\footnote{Remember that $\widetilde \phi$, the Legendre transform variable, is not the same as $\phi$, the integration variable in Eq.~(\ref{22}).  $\widetilde \phi$ is sometimes written as $\phi_{\rm cl}(x)$, and called the classical (or background)  field.}

\begin{align}
\Gamma[\widetilde \phi] &= W[J]- J \widetilde \phi, & \widetilde \phi &= \frac{\delta W}{\delta J}.
\label{23}
\end{align}

Quantum corrections to the theory are given by computing $\Gamma[\widetilde \phi]$.  At one-loop,
\begin{align}
\Gamma[\widetilde \phi] &= S[\widetilde \phi] + \frac i2 \ln \det \left( \frac{\delta^2 S}{\delta \phi^\mfi \delta \phi^\mfj} \right)_{\phi=\widetilde \phi}
\,.
\label{24}
\end{align}
From Eq.~(\ref{23}), we see that
\begin{align}
\widetilde \phi &= \frac{\delta W}{\delta J} = \vev{\phi}
\end{align}
is the
the vacuum expectation value of $\phi$. Eq.~(\ref{24}) is equivalent to summing the one-loop diagrams for $S[\phi]$. For example, the one-loop Coleman-Weinberg effective potential, which was originally derived by summing one-loop graphs, is equal to the determinant in Eq.~(\ref{24}) evaluated for a constant background field $\widetilde \phi$. The variation of the action is computed using
\begin{align}
\phi = \widetilde \phi+\eta
\label{27a}
\end{align}
and expanding in $\eta$.

Green's functions for the sigma model are generated by the action $S[\varphi]$, which is the spacetime integral of the Lagrangian
\begin{align}
\mathscr{L} &= \frac12 g_{ab} (\varphi) D_\mu \varphi^a D^\mu \varphi^b + J_a \varphi^a,
\label{27}
\end{align}
where we have included gauge
interactions by using the gauge covariant derivative $D_\mu$. 
$\Gamma[\widetilde \varphi]$ is given by taking the Legendre transform with respect to $J$. Expanding about $\widetilde \varphi$ gives the variation of the action
\begin{align}
\delta S &=  - \int \rd x \ g_{ab} (\widetilde \varphi) \eta^a\left[ \MsD^\mu (D_\mu \widetilde \varphi) \right]^b\, ,
\label{29}
\end{align}
where
\begin{align}\label{EOMCurv}
 \MsD_\mu\equiv&\, \partial_\mu \delta^a_b +\McD_b \left(D^\mu \varphi^a\right), 
 & \McD_a V^b&=\frac{\partial V^b}{\partial \varphi^a}+\Gamma^{b}_{ac}(\widetilde \varphi)V^c,
\end{align}
is the covariant derivative on the scalar manifold ${\cal G}/{\cal H} = S^3$, $V^b$ is a general vector, and $\Gamma^a_{bc}(\widetilde \varphi)$ is the Christoffel symbol constructed using the metric $g_{ab}(\widetilde \varphi)$ in Eq.~(\ref{27}). It is important to remember that the metric and tensors are in \emph{scalar field space}, which is curved.  Spacetime is flat in our analysis. The indices $a,b,c$, etc. refer to the  scalar manifold $S^3$, and the covariant derivatives are in scalar field space. 

The second variation of the action is
\begin{align}
\delta^2 S &= \frac12 \int \rd x \biggl[ g_{ab} (\MsD_\mu \eta)^a (\MsD^\mu \eta)^b-R_{abcd}D_\mu \widetilde \varphi^a D^\mu \widetilde \varphi^c \eta^b \eta^d \nn 
&- g_{af} \Gamma^f_{bc} \eta^b \eta^c(\MsD^\mu D_\mu\widetilde \varphi)^a \biggr] .
\label{30}
\end{align}
The first two terms in Eq.~(\ref{30}) are covariant, but the last term is not.  This last term results in the non-covariant terms found in one-loop calculations in the literature~\cite{Appelquist:1980ae,Gavela:2014uta}.

The origin of the non-covariant terms can be traced back to the expansion of $S[\phi + \eta]$ in $\eta$ to compute the second variation. Under a coordinate transformation $\varphi^{\prime\,a} = f^a(\varphi)$, $(\varphi^\prime + \eta^\prime)^a = f^a (\varphi+\eta) $
so that
\begin{align}
\eta^{\prime\,a} 
& = \frac{\partial \varphi^{\prime\,a} }{\partial \varphi^b} \eta^b
+ \frac12 \frac{\partial^2 \varphi^{\prime\,a} }{\partial \varphi^b \partial \varphi^c} \eta^b \eta^c + \ldots
\end{align}
The relation between $\eta^\prime$ and $\eta$ is the transformation rule for a tensor to
first order in $\eta$, but not to second order. Thus, $\delta S$ is covariant, but $\delta^2 S$ is not. To remedy the problem, we can relate $\widetilde \varphi+\delta \varphi$ and $\widetilde \varphi$ in a covariant way using geodesics on the curved manifold~\cite{Honerkamp:1971sh}. Pick $\zeta^a$ to be the tangent velocity vector of the geodesic with parameter $\lambda$, which starts at $\widetilde\varphi$ with $\lambda=0$ and ends at $\widetilde \varphi+\delta \varphi$ with $\lambda=1$. The geodesic equation is
\begin{align}
\frac{\rd^2 \varphi^a}{\rd \lambda^2} + \Gamma^a_{bc} \frac{\rd \varphi^b}{\rd \lambda}\frac{\rd \varphi^c}{\rd \lambda} &=0,
\end{align}
with the power series solution
\begin{align}
\varphi^a(\lambda) &= \widetilde \varphi^a + \zeta^a \lambda  - \frac12  \Gamma^a_{bc}\zeta^b \zeta^c \lambda^2  + \ldots
\end{align}
Using the expansion
\begin{align}
\varphi^a &= \widetilde \varphi^a + \zeta^a   - \frac12  \Gamma^a_{bc}(\widetilde \varphi) \zeta^b \zeta^c  + \ldots
\label{27b}
\end{align}
and expanding in $\zeta$, which are geodesic (normal) coordinates, changes the functional derivatives of the action into covariant functional derivatives,
\begin{align}
\tilde \McD_a S=&\frac{\delta S}{\delta \varphi^a}\,,&
\tilde \McD_b\tilde \McD_a S=&\frac{\delta^2 S}{\delta \varphi^b\delta \varphi^a}-\Gamma^c_{ab}\frac{\delta S}{\delta \varphi^c}\,.
\label{37}
\end{align}
The first variation is the same as before, Eq.~(\ref{29}), with $\eta \to \zeta$, but the second variation is modified to
\begin{align}
\delta^2 S &= \frac12 \int \rd x \biggl[ g_{ab} (\MsD_\mu \zeta)^a (\MsD^\mu \zeta)^b-R_{a b c d} D_\mu \widetilde \varphi^a D^\mu \widetilde \varphi^c \zeta^b \zeta^d  \biggr],
\label{30a}
\end{align}
which is covariant. The difference between Eqs.~(\ref{27a}) and~(\ref{27b}) is a field redefinition, so that the second variation of the action $\delta^2 S$ changes by an equation of motion term $\Gamma^c_{ab} {\delta S}/{\delta \varphi^c}$. This observation explains why the non-covariant terms found using Eq.~(\ref{30}) vanish on-shell; they can be eliminated by a field redefinition which does not change the $S$-matrix.

The above formalism can be applied to HEFT, with scalar fields $\phi^{\mfi}=\{\varphi^a,h\}$.  The order $p^2$ HEFT Lagrangian is 
\begin{align}
\mathscr L=\frac12g_{\mfi\mfj}(\phi)\ D_\mu\phi^{\mfi} D^\mu \phi^{\mfj}+\mathcal I(\phi),
\label{p2Lag}
\end{align}
where $\mathcal{I}(\phi)$ is a ${\cal G}$-invariant potential, the metric is given in Eq.~(\ref{GMtrc}), and 
the gauge covariant derivative is given in Eqs.~(\ref{GgCovCurv}--\ref{GenCurv}).
In geometrical terms, $t_A^\mathfrak{i}$ are Killing vectors that generate the gauge symmetry ${\cal G}_{\rm gauge}= SU(2)_L\times U(1)_Y\subset {\cal G}$ 
of the scalar manifold ${\cal M}$.

${\cal G}$-invariance of the Lagrangian requires
\begin{align}
\mathcal{L}_{t_A}(g_{\mfi\mfj}) = &t^\mfk_A\McD_\mfk g_{\mfi\mfj}+g_{\mfk\mfj}\McD_\mfi t^\mfk_A + g_{\mfi\mfk}\McD_\mfj t^\mfk_A =0,\nn
   \mathcal{L}_{t_A}(\mathcal I)=& t^\mfj_A \McD_\mfj {\cal I} =0,
\end{align}
where $\mathcal{L}$ is the Lie derivative and $A$ runs through weak isospin and hypercharge
generators $A=1,..,4$. The generators satisfy the commutation relations,
\begin{align}
\left[t_A, t_B\right]^\mathfrak{i} &=\mathcal{L}_{t_A}(t_B^\mfi)= f_{AB}^C t_C^\mathfrak{i}
\end{align}
where $\left[*,*\right]$ is the Lie bracket, and $f_{AB}^C$ are the structure constants of the Lie algebra $\mathfrak{g}$ of $G$.

The classical equation of motion is:
\begin{align}
\tilde \McD_\mathfrak{i} S =\frac{\delta S}{\delta \phi^\mathfrak{i}}=-\MsD_\mu (D^\mu\phi)_\mathfrak{i}+\McD_\mathfrak{i}\mathcal I=0 \label{EOMexpl}
\end{align}
with $\MsD_\mu=\delta^\mathfrak{i}_{\,\mathfrak{j}}\partial_\mu+\McD_\mathfrak{j}(D_\mu \phi)^\mathfrak{i}$ as given in Eq.~(\ref{EOMCurv}).

The one-loop correction to a general Lagrangian of the form Eq.~(\ref{p2Lag}) has been computed by 't~Hooft~\cite{tHooft:1973us},
and is given in $4-2\epsilon$ dimensions by
\begin{align}
\Delta \Gamma &=
\frac i2\log \det\left(-\tilde \McD_\mfi\tilde \McD_\mfj \mathcal L\right) \nn
&= \frac{1}{16\pi^2 \epsilon} \int \rd x\, \mbox{Tr} \left\{\frac{\Gamma_{\mu\nu}^2}{24}+\frac{X^2}{4} \right\},
\label{43}
\end{align}
where the antisymmetric tensor $\Gamma_{\mu\nu}$ and scalar $X$ are~\cite{p2}:
\begin{align}
\left[\Gamma_{\mu\nu} \right]^\mathfrak{i}{}_\mathfrak{j} &=\left[\mathscr{D}_\mu\,,\mathscr{D}_\nu\right]=R^\mathfrak{i}_\mathfrak{\,jkl}D^\mu\phi^\mathfrak{k}
D^\nu\phi^\mathfrak{l}+ \left(\mathscr{A}^{\mu\nu}\right)^\mathfrak{i}_{\,\mathfrak{j}}, \nn
X^\mathfrak{i}{}_\mathfrak{j} &=\McD^\mathfrak{i} \McD_\mathfrak{j}\mathcal I-R^\mathfrak{i}_\mathfrak{\,kjl}D^\mu\phi^\mathfrak{k} D_\mu\phi^\mathfrak{l}, \nn
\left(\mathscr{A}^{\mu\nu}\right)^\mathfrak{i}_\mathfrak{\,j} &=\left(\partial_{[\mu} A_{\nu]}^B+f^B_{CD}A_\mu^CA_\nu^D\right)\McD_\mfj t^\mathfrak{i}_B,
\label{GrandFS}
\end{align}
and $R^\mathfrak{i}_\mathfrak{\,jkl}(\phi)$ is the Riemann tensor of $\mathcal M$. The  curvature of scalar field space
is multiplied by terms with two derivatives of $\phi$, so that Eq.~(\ref{43}) generates $O(p^4)$ terms at one-loop
if $\mathcal{M}$ is not flat. The divergent contribution Eq.~(\ref{43}) is cancelled by the one-loop counterterms.

\bigskip
\section{Renormalization of HEFT}

We return to the HEFT Lagrangian from Sec.~\ref{sec:curvature},
\begin{align}
\mathscr L&=\frac12\partial_\mu h\,\partial^\mu h +\frac 12 F(h)^2 g_{ab}(\varphi) D_\mu \varphi^a D_\mu \varphi^b\nn
& -V(h)+K(h)\, w^{{i}} u^{{i}} (\varphi),
\label{45}
\end{align}
where $D_\mu$ is the gauge 
covariant derivative from Eqs.~(\ref{GgCovCurv}--\ref{GenCurv}), and we have added two additional terms to the Lagrangian. $V(h)$ is an arbitrary Higgs potential.
The second additional term is given in terms of an arbitrary function $K(h)$,
the $4$-dimensional unit vector $u^{{i}}(\varphi)$ in Eq.~(\ref{udef}), and the $4$-dimensional vector $w^{{i}}$ constructed from fermion scalar bilinears
\begin{align}
w^{{i}} =&\bar q_L  \sigma^{{i}}  \,Y_q\, q_R+\bar \ell_L  \sigma^{{i}} \,Y_\ell\, \ell_R+ \text{h.c.},
\end{align}
where $\sigma^{{i}}=\{i \bm{\sigma}, \openone\}$, $q_L$ and $\ell_L$ are the left-handed quark and lepton doublets, $q_R=(u_R\,,d_R)$ and $\ell_R=(\nu_R\,,e_R)$ are
the right-handed quark and lepton doublets, and $Y_q= \mbox{diag} (Y_u\,,Y_d)$ and $Y_\ell= \mbox{diag} (Y_\nu\,,Y_e)$ are the quark and lepton Yukawa couplings.
The right-handed neutrino field $\nu_R$ drops out of the Lagrangian if $Y_\nu=0$.  In our choice of the above Lagrangian, we are assuming that fermion masses are generated from the Yukawa coupling matrices after spontaneous symmetry breakdown.

The one-loop divergent contribution from scalar loops can be computed from Eq.~(\ref{GrandFS}) to be $\Delta \Gamma = 1/(32\pi^2 \epsilon)Z$,
\begin{widetext}
\begin{align}
Z &=  \frac12 \left( V^{\prime \prime} - K^{\prime \prime}  w \cdot u\right)^2 + \left( \left( K/(v F) \right)^\prime \right)^2 \left[ w \cdot w - (w \cdot u)^2 \right] + \frac12 N_\varphi\left[ \left( \frac{F^{\prime \prime}}{F}  \right) (\partial_\mu h \partial^\mu h) - \frac{V^\prime F^\prime}{F} + (w \cdot u) \left( \frac{F^\prime K^\prime}{F}-\frac{K}{v^2 F^2} \right) \right]^2 \nn
& -\left[ (v^2 F F^{\prime\prime}) \left( V^{\prime \prime} - K^{\prime \prime}  u \cdot w \right) + (N_\varphi-1)\left[1-(v F^\prime)^2\right]
\left\{ - \frac{V^\prime F^\prime}{F} + (w \cdot u) \left( \frac{F^\prime K^\prime}{F}-\frac{K}{v^2 F^2} \right) \right\}  \right]  (D_\mu u \cdot D^\mu u)  \nn
& -\left[ \frac13 (v F^{\prime \prime})^2  + (N_\varphi-1) \left[1-(v F^\prime)^2\right]\frac{F^{\prime\prime}}{F}  \right]  (\partial_\nu h \partial^\nu h)   (D_\mu u \cdot D^\mu u)  + \frac23\left[1-(v F^\prime)^2\right]^2 (D_\mu u \cdot D_\nu u)^2 \nn
&+ \left[ \frac12(v^2 F F^{\prime\prime})^2  +\frac{3N_\varphi-7}{6} \left[1-(v F^\prime)^2\right]^2  \right] (D_\mu u \cdot D^\mu u)^2  + \frac43 (v F^{\prime\prime})^2 (\partial^\mu h \partial^\nu h)  (D_\mu u \cdot D_\nu u)  - 2 F^{\prime\prime} (\partial^\mu h) \left(K/F \right)^\prime ( w \cdot D_\mu u) \nn
&
- \frac13  \left[1-(v F^\prime)^2\right]  (D^\mu u)^T A_{\mu \nu} (D^\nu u)  -\frac23 (vF^{\prime })(vF^{\prime \prime}) (\partial_\mu h) (D_\nu u)^T A^{\mu \nu} u+\frac{1}{12} \text{tr} (A_{\mu \nu}  A^{\mu \nu}  ) + \frac 16 \left[(v F^{\prime })^2 -1 \right] u^T (A_{\mu \nu}  A^{\mu \nu}  ) u ,
\label{CT2}
\end{align}
\end{widetext}
where  the gauge covariant derivative $D_\mu$ is defined in Eq.~(\ref{GgCovCurv}) and the gauge field strength $A_{\mu\nu}\equiv\left[D_\mu,D_\nu\right]$.  
The above expression gives the one-loop divergent contribution from scalar loops in HEFT with scalars in the internal loops, and respects the underlying ${\cal G}=O(4)$ symmetry of the theory. This expression agrees with the one-loop result computed previously in Ref.~\cite{Guo:2015isa}.

The expression Eq.~(\ref{CT2}) of the one-loop divergences of HEFT can be used to analyze several interesting special cases in a unified way.
\begin{itemize}
\item  {\bf Chiral Perturbation Theory:} If $F(h)$ and $K(h)$ are constants, we get the one-loop correction to the non-linear sigma model. This special case includes the well known results of chiral perturbation theory.

\item {\bf The SM Higgs boson: } If $F(h)=K(h)=v+h$ and $V(h)= (\lambda/4)(h^2+2hv)^2$, the HEFT Lagrangian reduces to the SM Higgs Lagrangian.  In this case, the first derivatives 
$F^\prime(h)=K^\prime(h)=1$, the second derivatives $F^{\prime\prime}(h)=K^{\prime\prime}(h)=0$, and, as noted in Sec.~\ref{sec:curvature}, the curvature 
tensor $R^\mfi_{\mfj\mfk\mfl}(\phi_H)$ vanishes.  As a result,
all order $p^4$ terms disappear, and the theory is renormalizable, even though in the field parametrization chosen here, renormalizability is not obvious. 

\item {\bf The Higgs  as a Goldstone boson: } Assume the Higgs boson $h$ and ``eaten" scalars $\varphi^a$ all are Goldstone bosons resulting from dynamical breaking of a global symmetry at a high energy scale ~\cite{Dugan:1984hq,Kaplan:1983fs}.  As a concrete example, consider the symmetry breaking pattern ${\cal G}/{\cal H}=O(5)/O(4)$~\cite{Agashe:2004rs}.  Then, the scalar theory can be described in terms of a unit vector $U$ in $5$ dimensions, e.g. $U=\left(\cos(h/f),\sin(h/f)\,u^{{i}}(\varphi)\right)$ 
where $u^{{i}}(\varphi)$ is the $4$-dimensional unit vector of Eq.~(\ref{udef}) and $F(h)=\sin(h/f)$.  One can check that
the higher-dimensional operators reduce to a number of invariants constructed with $u$, multiplied by calculable singlet functions of $h$ which have been 
computed in Ref.~\cite{Alonso:2014wta}.  Although the theory is not renormalizable in this special case, the reduction in number of parameters is drastic: instead of
arbitrary functions of $h$ multiplying each operator, one obtains calculable functions of $h$ which are determined up to an overall normalization coefficient.

\item  {\bf The Higgs boson as a dilaton: }A Higgs boson which is the Goldstone boson of spontaneously broken scale invariance 
closely resembles the SM Higgs boson, see e.g.~\cite{Goldberger:2008zz}.  The Lagrangian for the dilaton can be obtained by restoring scale invariance 
by the substitution $v \rightarrow v e^{\tau/v}$, where $\tau$ is the dilaton field, which transforms by a shift under scaling.  This prescription yields a kinetic term, 
$e^{2\tau/v}(\partial \tau)^2/2$, which is not normalized as in Eq.~(\ref{GMtrc}), so a variable change $h/v=e^{\tau/v}-1$ is needed to compare with our study.
In terms of the correctly normalized field $h$, one obtains $F(h)=K(h)=v+h$ and a vanishing curvature, like the SM Higgs case.  In particular,
the $p^4$ dilaton-dilaton scattering term vanishes, which is related to the $a$-theorem~\cite{Komargodski:2011vj}.
Note that for a non-vanishing scattering, the running would induce an increasing coefficient for decreasing energy, which is a contradiction of the $a$-theorem.
Finally, to preserve scale invariance at one loop, dimensional regularization must be modified by the substitution $\mu \to \mu \, e^{\tau/v}$,
which implies that the dilaton couplings are proportional to anomalous dimensions.  This property sets the dilaton Higgs boson apart from the SM Higgs boson.

\end{itemize}

Finally, we can make a connection with the non-invariant terms found in Refs.~\cite{Appelquist:1980ae,Gavela:2014uta}.
In ordinary perturbation theory, one gets the second variation of the action Eq.~(\ref{30}), which contains a non-covariant term. At one loop, this will translate into extra terms in the Lagrangian
\begin{align}\label{Ninv}
\delta \mathscr L=\frac{1}{32\pi^2\epsilon}\left(X^{\mfi\mfj}\Gamma^\mfk_{\mfi\mfj}\frac{\delta S}{\delta \phi^\mfk}+\frac12\Gamma^\mfk_{\mfi\mfj}\frac{\delta S}{\delta \phi^\mfk}\Gamma^{\mfl\mfi\mfj}\frac{\delta S}{\delta \phi^\mfl}\right),
\end{align}
which are not invariant under ${\cal G}$ and are precisely the set of terms found in \cite{Appelquist:1980ae,Gavela:2014uta}. 
Consider an explicit example.  Ref.~\cite{Gavela:2014uta} uses the metric
\begin{align}\label{gmetric}
g_{ab}(\varphi) &= \left[ \delta_{ab}\left(1+2\,\eta \, \frac{{\bm \varphi}^2}{v^2}\right) + \frac{ \varphi_a \varphi_b}{v^2}
\left(1+4\,\eta\right) \right] +\mathcal{O}{(\varphi^4)},
\end{align}
where $\eta$ is a variable that characterizes different parametrizations. The second term in Eq.~(\ref{Ninv})
depends only on the geometry, and it can be computed using the metric Eq.~(\ref{gmetric}):
\begin{align}
\Gamma^\mfi_{\mfk\mfl}\Gamma^{\mfj\mfk\mfl}=
\left[ \begin{array}{cc}N_\varphi\frac{(F^\prime)^2}{F^2}&-\frac{F^\prime \varphi^b}{v^2F^3}C_\varphi \\
-\frac{F^\prime \varphi^a}{v^2F^3}C_\varphi&2\frac{(F^\prime)^2}{v^2F^4}g^{ab}+\frac{1}{F^4} \Gamma^a_{cd}\Gamma^{bcd} \end{array}\right],
\end{align}
where 
\begin{align}
C_\varphi &\equiv 2\eta(N_\varphi+2)+N_\varphi , \\
v^4\Gamma^a_{cd}\Gamma^{bcd}&=8\eta^2\delta^{ab}(\bm{\varphi \cdot \varphi}) \nn \nonumber
&+\varphi^a\varphi^b\left[ N_\varphi(1+4\eta+4\eta^2)+8\eta+24\eta^2\right].
\end{align}
For example, two of the operators generated are
\begin{align*}
\frac{1}{32\pi^2\epsilon}\!\left\{\!\left(\frac32+10\eta +18\eta^2\right)\!\frac{\left({\bm\varphi}\Box {\bm\varphi}\right)^2}{v^4}-c_1\!\left(3+10\eta\right)\!\frac{{\bm\varphi}\Box {\bm\varphi}\Box h}{v^3} \right\},
\end{align*}
in agreement with Ref.~\cite{Gavela:2014uta}.

\section{Conclusions}
 
 In conclusion, the geometric formulation of HEFT makes clear that physical observables of HEFT are given by the geometric invariants of the scalar field manifold, which are independent of the scalar fields used to parametrize the Lagrangian.  From this perspective, whether the electroweak gauge symmetry breaking is parametrized linearly or nonlinearly is not a physical distinction, and the SM Higgs sector in a non-linear realization is still renormalizable.  Instead, the physical distinction to be made is whether the scalar field manifold is flat or curved.  The scalar sector of HEFT allows for the possibility that the scalar curvature does not vanish.  We have given several examples of how the scalar curvature can be directly measured.  Future precision measurements of Higgs boson couplings and cross-sections can be used to constrain the curvature, and hence the scale of new physics.   

\acknowledgments

AM would like to thank Luis Alvarez-Gaum\'e for helpful discussion on non-linear sigma models. 

This work was supported in part by DOE grant DE-SC0009919, Spanish MINECO's ``Centro de Excelencia Severo Ochoa'' Programme under grant SEV-2012-0249, and by grants from the Simons Foundation (\#340282 to Elizabeth Jenkins and \#340281 to Aneesh Manohar).

\bibliography{Higgs}

\begin{thebibliography}{24}
\expandafter\ifx\csname natexlab\endcsname\relax\def\natexlab#1{#1}\fi
\expandafter\ifx\csname bibnamefont\endcsname\relax
  \def\bibnamefont#1{#1}\fi
\expandafter\ifx\csname bibfnamefont\endcsname\relax
  \def\bibfnamefont#1{#1}\fi
\expandafter\ifx\csname citenamefont\endcsname\relax
  \def\citenamefont#1{#1}\fi
\expandafter\ifx\csname url\endcsname\relax
  \def\url#1{\texttt{#1}}\fi
\expandafter\ifx\csname urlprefix\endcsname\relax\def\urlprefix{URL }\fi
\providecommand{\bibinfo}[2]{#2}
\providecommand{\eprint}[2][]{\url{#2}}

\bibitem[{\citenamefont{Feruglio}(1993)}]{Feruglio:1992wf}
\bibinfo{author}{\bibfnamefont{F.}~\bibnamefont{Feruglio}},
  \bibinfo{journal}{Int. J. Mod. Phys.} \textbf{\bibinfo{volume}{A8}},
  \bibinfo{pages}{4937} (\bibinfo{year}{1993}), \eprint{hep-ph/9301281}.

\bibitem[{\citenamefont{Grinstein and Trott}(2007)}]{Grinstein:2007iv}
\bibinfo{author}{\bibfnamefont{B.}~\bibnamefont{Grinstein}} \bibnamefont{and}
  \bibinfo{author}{\bibfnamefont{M.}~\bibnamefont{Trott}},
  \bibinfo{journal}{Phys. Rev.} \textbf{\bibinfo{volume}{D76}},
  \bibinfo{pages}{073002} (\bibinfo{year}{2007}), \eprint{0704.1505}.

\bibitem[{\citenamefont{Coleman et~al.}(1969)\citenamefont{Coleman, Wess, and
  Zumino}}]{Coleman:1969sm}
\bibinfo{author}{\bibfnamefont{S.~R.} \bibnamefont{Coleman}},
  \bibinfo{author}{\bibfnamefont{J.}~\bibnamefont{Wess}}, \bibnamefont{and}
  \bibinfo{author}{\bibfnamefont{B.}~\bibnamefont{Zumino}},
  \bibinfo{journal}{Phys. Rev.} \textbf{\bibinfo{volume}{177}},
  \bibinfo{pages}{2239} (\bibinfo{year}{1969}).

\bibitem[{\citenamefont{Callan et~al.}(1969)\citenamefont{Callan, Coleman,
  Wess, and Zumino}}]{Callan:1969sn}
\bibinfo{author}{\bibfnamefont{C.~G.} \bibnamefont{Callan}},
  \bibinfo{author}{\bibfnamefont{S.~R.} \bibnamefont{Coleman}},
  \bibinfo{author}{\bibfnamefont{J.}~\bibnamefont{Wess}}, \bibnamefont{and}
  \bibinfo{author}{\bibfnamefont{B.}~\bibnamefont{Zumino}},
  \bibinfo{journal}{Phys. Rev.} \textbf{\bibinfo{volume}{177}},
  \bibinfo{pages}{2247} (\bibinfo{year}{1969}).

\bibitem[{\citenamefont{Honerkamp}(1972)}]{Honerkamp:1971sh}
\bibinfo{author}{\bibfnamefont{J.}~\bibnamefont{Honerkamp}},
  \bibinfo{journal}{Nucl. Phys.} \textbf{\bibinfo{volume}{B36}},
  \bibinfo{pages}{130} (\bibinfo{year}{1972}).

\bibitem[{\citenamefont{Tataru}(1975)}]{Tataru:1975ys}
\bibinfo{author}{\bibfnamefont{L.}~\bibnamefont{Tataru}},
  \bibinfo{journal}{Phys. Rev.} \textbf{\bibinfo{volume}{D12}},
  \bibinfo{pages}{3351} (\bibinfo{year}{1975}).

\bibitem[{\citenamefont{Alvarez-Gaume et~al.}(1981)\citenamefont{Alvarez-Gaume,
  Freedman, and Mukhi}}]{AlvarezGaume:1981hn}
\bibinfo{author}{\bibfnamefont{L.}~\bibnamefont{Alvarez-Gaume}},
  \bibinfo{author}{\bibfnamefont{D.~Z.} \bibnamefont{Freedman}},
  \bibnamefont{and} \bibinfo{author}{\bibfnamefont{S.}~\bibnamefont{Mukhi}},
  \bibinfo{journal}{Annals Phys.} \textbf{\bibinfo{volume}{134}},
  \bibinfo{pages}{85} (\bibinfo{year}{1981}).

\bibitem[{\citenamefont{Alvarez-Gaume and
  Freedman}(1981)}]{AlvarezGaume:1981hm}
\bibinfo{author}{\bibfnamefont{L.}~\bibnamefont{Alvarez-Gaume}}
  \bibnamefont{and} \bibinfo{author}{\bibfnamefont{D.~Z.}
  \bibnamefont{Freedman}}, \bibinfo{journal}{Commun. Math. Phys.}
  \textbf{\bibinfo{volume}{80}}, \bibinfo{pages}{443} (\bibinfo{year}{1981}).

\bibitem[{\citenamefont{Gaillard}(1986)}]{Gaillard:1985uh}
\bibinfo{author}{\bibfnamefont{M.~K.} \bibnamefont{Gaillard}},
  \bibinfo{journal}{Nucl. Phys.} \textbf{\bibinfo{volume}{B268}},
  \bibinfo{pages}{669} (\bibinfo{year}{1986}).

\bibitem[{\citenamefont{Alonso et~al.}(2016)\citenamefont{Alonso, Jenkins, and
  Manohar}}]{p2}
\bibinfo{author}{\bibfnamefont{R.}~\bibnamefont{Alonso}},
  \bibinfo{author}{\bibfnamefont{E.~E.} \bibnamefont{Jenkins}},
  \bibnamefont{and} \bibinfo{author}{\bibfnamefont{A.~V.}
  \bibnamefont{Manohar}} (\bibinfo{year}{2016}).

\bibitem[{\citenamefont{Appelquist and Bernard}(1980)}]{Appelquist:1980vg}
\bibinfo{author}{\bibfnamefont{T.}~\bibnamefont{Appelquist}} \bibnamefont{and}
  \bibinfo{author}{\bibfnamefont{C.~W.} \bibnamefont{Bernard}},
  \bibinfo{journal}{Phys. Rev.} \textbf{\bibinfo{volume}{D22}},
  \bibinfo{pages}{200} (\bibinfo{year}{1980}).

\bibitem[{\citenamefont{Appelquist and Bernard}(1981)}]{Appelquist:1980ae}
\bibinfo{author}{\bibfnamefont{T.}~\bibnamefont{Appelquist}} \bibnamefont{and}
  \bibinfo{author}{\bibfnamefont{C.~W.} \bibnamefont{Bernard}},
  \bibinfo{journal}{Phys. Rev.} \textbf{\bibinfo{volume}{D23}},
  \bibinfo{pages}{425} (\bibinfo{year}{1981}).

\bibitem[{\citenamefont{Gavela et~al.}(2015)\citenamefont{Gavela, Kanshin,
  Machado, and Saa}}]{Gavela:2014uta}
\bibinfo{author}{\bibfnamefont{M.}~\bibnamefont{Gavela}},
  \bibinfo{author}{\bibfnamefont{K.}~\bibnamefont{Kanshin}},
  \bibinfo{author}{\bibfnamefont{P.}~\bibnamefont{Machado}}, \bibnamefont{and}
  \bibinfo{author}{\bibfnamefont{S.}~\bibnamefont{Saa}},
  \bibinfo{journal}{JHEP} \textbf{\bibinfo{volume}{1503}}, \bibinfo{pages}{043}
  (\bibinfo{year}{2015}), \eprint{1409.1571}.

\bibitem[{\citenamefont{Dugan et~al.}(1985)\citenamefont{Dugan, Georgi, and
  Kaplan}}]{Dugan:1984hq}
\bibinfo{author}{\bibfnamefont{M.~J.} \bibnamefont{Dugan}},
  \bibinfo{author}{\bibfnamefont{H.}~\bibnamefont{Georgi}}, \bibnamefont{and}
  \bibinfo{author}{\bibfnamefont{D.~B.} \bibnamefont{Kaplan}},
  \bibinfo{journal}{Nucl. Phys.} \textbf{\bibinfo{volume}{B254}},
  \bibinfo{pages}{299} (\bibinfo{year}{1985}).

\bibitem[{\citenamefont{Kaplan and Georgi}(1984)}]{Kaplan:1983fs}
\bibinfo{author}{\bibfnamefont{D.~B.} \bibnamefont{Kaplan}} \bibnamefont{and}
  \bibinfo{author}{\bibfnamefont{H.}~\bibnamefont{Georgi}},
  \bibinfo{journal}{Phys. Lett.} \textbf{\bibinfo{volume}{B136}},
  \bibinfo{pages}{183} (\bibinfo{year}{1984}).

\bibitem[{\citenamefont{Peskin and Takeuchi}(1990)}]{Peskin:1990zt}
\bibinfo{author}{\bibfnamefont{M.~E.} \bibnamefont{Peskin}} \bibnamefont{and}
  \bibinfo{author}{\bibfnamefont{T.}~\bibnamefont{Takeuchi}},
  \bibinfo{journal}{Phys. Rev. Lett.} \textbf{\bibinfo{volume}{65}},
  \bibinfo{pages}{964} (\bibinfo{year}{1990}).

\bibitem[{\citenamefont{Barbieri et~al.}(2007)\citenamefont{Barbieri,
  Bellazzini, Rychkov, and Varagnolo}}]{Barbieri:2007bh}
\bibinfo{author}{\bibfnamefont{R.}~\bibnamefont{Barbieri}},
  \bibinfo{author}{\bibfnamefont{B.}~\bibnamefont{Bellazzini}},
  \bibinfo{author}{\bibfnamefont{V.~S.} \bibnamefont{Rychkov}},
  \bibnamefont{and}
  \bibinfo{author}{\bibfnamefont{A.}~\bibnamefont{Varagnolo}},
  \bibinfo{journal}{Phys. Rev.} \textbf{\bibinfo{volume}{D76}},
  \bibinfo{pages}{115008} (\bibinfo{year}{2007}), \eprint{0706.0432}.

\bibitem[{\citenamefont{Manohar and Georgi}(1984)}]{Manohar:1983md}
\bibinfo{author}{\bibfnamefont{A.}~\bibnamefont{Manohar}} \bibnamefont{and}
  \bibinfo{author}{\bibfnamefont{H.}~\bibnamefont{Georgi}},
  \bibinfo{journal}{Nucl. Phys.} \textbf{\bibinfo{volume}{B234}},
  \bibinfo{pages}{189} (\bibinfo{year}{1984}).

\bibitem[{\citenamefont{'t~Hooft}(1973)}]{tHooft:1973us}
\bibinfo{author}{\bibfnamefont{G.}~\bibnamefont{'t~Hooft}},
  \bibinfo{journal}{Nucl. Phys.} \textbf{\bibinfo{volume}{B62}},
  \bibinfo{pages}{444} (\bibinfo{year}{1973}).

\bibitem[{\citenamefont{Guo et~al.}(2015)\citenamefont{Guo, Ruiz-Femen{\'\i}a,
  and Sanz-Cillero}}]{Guo:2015isa}
\bibinfo{author}{\bibfnamefont{F.-K.} \bibnamefont{Guo}},
  \bibinfo{author}{\bibfnamefont{P.}~\bibnamefont{Ruiz-Femen{\'\i}a}},
  \bibnamefont{and} \bibinfo{author}{\bibfnamefont{J.~J.}
  \bibnamefont{Sanz-Cillero}}, \bibinfo{journal}{Phys. Rev.}
  \textbf{\bibinfo{volume}{D92}}, \bibinfo{pages}{074005}
  (\bibinfo{year}{2015}), \eprint{1506.04204}.

\bibitem[{\citenamefont{Agashe et~al.}(2005)\citenamefont{Agashe, Contino, and
  Pomarol}}]{Agashe:2004rs}
\bibinfo{author}{\bibfnamefont{K.}~\bibnamefont{Agashe}},
  \bibinfo{author}{\bibfnamefont{R.}~\bibnamefont{Contino}}, \bibnamefont{and}
  \bibinfo{author}{\bibfnamefont{A.}~\bibnamefont{Pomarol}},
  \bibinfo{journal}{Nucl. Phys.} \textbf{\bibinfo{volume}{B719}},
  \bibinfo{pages}{165} (\bibinfo{year}{2005}), \eprint{hep-ph/0412089}.

\bibitem[{\citenamefont{Alonso et~al.}(2014)\citenamefont{Alonso, Brivio,
  Gavela, Merlo, and Rigolin}}]{Alonso:2014wta}
\bibinfo{author}{\bibfnamefont{R.}~\bibnamefont{Alonso}},
  \bibinfo{author}{\bibfnamefont{I.}~\bibnamefont{Brivio}},
  \bibinfo{author}{\bibfnamefont{B.}~\bibnamefont{Gavela}},
  \bibinfo{author}{\bibfnamefont{L.}~\bibnamefont{Merlo}}, \bibnamefont{and}
  \bibinfo{author}{\bibfnamefont{S.}~\bibnamefont{Rigolin}},
  \bibinfo{journal}{JHEP} \textbf{\bibinfo{volume}{1412}}, \bibinfo{pages}{034}
  (\bibinfo{year}{2014}), \eprint{1409.1589}.

\bibitem[{\citenamefont{Goldberger et~al.}(2008)\citenamefont{Goldberger,
  Grinstein, and Skiba}}]{Goldberger:2008zz}
\bibinfo{author}{\bibfnamefont{W.~D.} \bibnamefont{Goldberger}},
  \bibinfo{author}{\bibfnamefont{B.}~\bibnamefont{Grinstein}},
  \bibnamefont{and} \bibinfo{author}{\bibfnamefont{W.}~\bibnamefont{Skiba}},
  \bibinfo{journal}{Phys. Rev. Lett.} \textbf{\bibinfo{volume}{100}},
  \bibinfo{pages}{111802} (\bibinfo{year}{2008}), \eprint{0708.1463}.

\bibitem[{\citenamefont{Komargodski and Schwimmer}(2011)}]{Komargodski:2011vj}
\bibinfo{author}{\bibfnamefont{Z.}~\bibnamefont{Komargodski}} \bibnamefont{and}
  \bibinfo{author}{\bibfnamefont{A.}~\bibnamefont{Schwimmer}},
  \bibinfo{journal}{JHEP} \textbf{\bibinfo{volume}{1112}}, \bibinfo{pages}{099}
  (\bibinfo{year}{2011}), \eprint{1107.3987}.

\end{thebibliography}

\end{document}